\documentclass[aps,prx,groupedaddress,twocolumn,showpacs,longbibliography,10pt]{revtex4-2}
\usepackage{amsmath}
\usepackage{amsthm, amssymb,bbold,slashed}    
\usepackage{braket}              
\usepackage{graphicx}
\usepackage{draft}
\usepackage{color}
\usepackage{times, verbatim}      
\usepackage{mathtools,tikz}
\usetikzlibrary{matrix}
\usepackage{natbib}
\usepackage{multirow}
\usepackage{hyperref}
\usepackage{subcaption}

\definecolor{colour1}{rgb}{0.368417, 0.506779, 0.709798}
\definecolor{colour2}{rgb}{0.880722, 0.611041, 0.142051}
\definecolor{colour3}{rgb}{0,1,1}
\definecolor{colour4}{rgb}{0,1,0}
\definecolor{colour5}{rgb}{1,1,0}

\usepackage{amsthm}

\hypersetup{
       colorlinks=true,
}

\makeatletter
\newsavebox{\@brx}
\newcommand{\llangle}[1][]{\savebox{\@brx}{\(\m@th{#1\langle}\)}%
  \mathopen{\copy\@brx\kern-0.5\wd\@brx\usebox{\@brx}}}
\newcommand{\rrangle}[1][]{\savebox{\@brx}{\(\m@th{#1\rangle}\)}%
  \mathclose{\copy\@brx\kern-0.5\wd\@brx\usebox{\@brx}}}
\makeatother

\newcommand*\LP{\mathop{}\!\mathbin\Box}

\begin{document}
 
\author{Oleksandr Diatlyk, Himanshu Khanchandani, Fedor K.~Popov, and Yifan Wang$^{1}$
}
\affiliation{$^{1}$ Center for Cosmology and Particle Physics, New York University, New York, NY 10003, USA}

\title{ 
Effective Field Theory of Conformal Boundaries 
}

\begin{abstract}
We introduce an effective field theory (EFT) for conformal impurity by considering a pair of transversely displaced impurities and integrating out modes with mass inversely proportional to the separation distance. This EFT captures the universal signature of the impurity seen by a heavy local operator.
We focus on the case of conformal boundaries and derive universal formulas from this EFT for the boundary structure constants at high energy. We point out that the more familiar thermal EFT  for conformal field theory is a special case of this EFT with distinguished conformal boundaries. We also derive, for general conformal impurities,  non-positivity and convexity-like constraints on the Casimir energy which determines the leading EFT coefficient.

\end{abstract}

\date{\today}


\maketitle

\section{Introduction and Summary}

Matter rarely appears in pure form in nature. Instead it generally contains impurities which create 
discontinuities and irregularities in an otherwise homogeneous material. One may be tempted to assume that their effects are suppressed. However even isolated impurities can drastically modify the physical behavior of the host at the macroscopic scale, especially when the underlying system supports long-wavelength fluctuations. The archetypal example is the Kondo effect, where magnetic impurities in a metallic host produce an anomalous dip in the resistivity \cite{1964PThPh..32...37K}. The theoretical challenge is to explain and extract features of such many-body phenomena with impurities which typically involve strong interactions and emergent collective modes.  

Conformal Field Theory (CFT) is a powerful nonperturbative framework to describe many-body systems at and near criticality, where critical exponents and correlation functions are completely determined by an algebra of primary local operators $\cO_i(x)$ of scaling dimension $\Delta_i$ and structure constants $C_{ijk}$. This operator algebra is further subject to stringent consistency conditions from conformal symmetry, unitarity and associativity of the operator-product-expansion (OPE). On the one hand, this opens the way to systematically constrain and solve specific CFTs by the bootstrap method (see \cite{Poland:2018epd} for a review). On the other hand, by exploiting these consistency conditions, we can also deduce universal properties of the operator algebra data. A famous example for the latter is the Cardy formula for $d=2$ CFTs, which gives a precise prediction for the density $\rho(\Delta)$ of high energy states in terms of the central charge $c$ \cite{Cardy:1986ie}. The Cardy formula was originally derived from the modular invariance of the $d=2$ CFT but can be equivalently obtained from a thermal effective field theory in one dimension lower \cite{Shaghoulian:2015kta,Shaghoulian:2015lcn,Benjamin:2023qsc,Benjamin:2024kdg}. This is because these high energy states dominate the behavior of the CFT at high temperature, which can be characterized alternatively by an effective field theory with local action $S_{\rm eff}(\B)$ from reduction on the Euclidean time circle ${\mathbb{S}}^1_\B$ where $\B$ is the inverse temperature \cite{Bhattacharyya:2007vs,Jensen:2012jh,Banerjee:2012iz},
\ie 
Z(\B)\equiv  \int d\Delta \rho(\Delta)e^{-\B \Delta } =e^{-S_{\rm EFT}(\B)} \,.
\fe 
This second perspective generalizes the Cardy formula immediately to CFTs in general dimensions \cite{Shaghoulian:2015kta,Shaghoulian:2015lcn,Benjamin:2023qsc,Benjamin:2024kdg}, 
where the entropy $\log \rho(\Delta)$ at high energy  is controlled by the thermal effective action $S_{\rm eff}(\B)$
which admits a derivative expansion (suppressed by $\B$)
\ie 
S_{\rm EFT}(\B)={1\over \B^{d-1}}\int d^{d-1}x\sqrt{g} \left( -f + \dots \right) \,,
\label{thermaleff}
\fe 
where the leading Wilson coefficient $f$ is the negative of thermal free energy density,  which is proportional to $c$ in $d=2$.

In this work we introduce a generalization of the thermal effective theory reviewed above that will capture the universal behavior of new CFT data in the presence of impurities.
Impurities in critical systems are generally described by conformal defects at long distance. In particular, the Kondo effect is elegantly explained by viewing the impurity as a nontrivial boundary condition for the $d=2$ free fermion CFT that models the electron gas upon an s-wave reduction \cite{affleck1990current, affleck1991critical, affleck1991kondo, Affleck:1995ge}. The nontrivial conformal boundary that emerges at long distances captures the physical signatures of the Kondo impurity. In general, conformal defects are characterized by the one-point function of bulk local operators, which no longer vanish due to the defect insertion and introduce new structure constants for the combined system, known as defect CFT (DCFT) \cite{Cardy:1989ir,Cardy:1991tv,Billo:2016cpy}. For example, the one-point function of a scalar primary operator $\cO_i$ of weight $\Delta_i$ in the presence of a flat conformal boundary $\cB$ takes the following form with constant coefficient $C_i$,
\ie 
\la \cO_i (x)\ra_\cB ={C_i\over 2^{\Delta_i}|x_\perp|^{\Delta_i}}\,,
\label{Bopf}
\fe 
where $x_\perp$ is the perpendicular distance to $\cB$ and spinning primaries have vanishing one-point functions.
Defect data ($e.g.$ $C_i$ above) are subject to further bootstrap-type constraints from the consistency of defect observables, such as the associativity of two-point function of bulk local operators which may factor through either a bulk or a defect OPE channel \cite{Cardy:1991tv,Billo:2016cpy}. The main goal of this work is derive universal behavior of such defect data by exploiting a novel effective field theory (EFT) approach,
which we refer to as the defect EFT (DEFT), 
that generalizes the more familiar Cardy formula for CFT data in the absence of defects. Our approach is  applicable to defects of general codimensions. We focus on the case of conformal boundaries in this letter to illustrate the main ideas and a more exhaustive analysis is contained in \cite{IshibashiDefLong}. One of the main results is a Cardy-like formula for the behavior of the structure constant $C_i$ in \eqref{Bopf} for operators of high dimensions.

The natural setup where the DEFT arises involves a pair of parallel planar defects $\cD_1(0)$ and $\cD_2(\D)$ in flat spacetime with transverse separation $\D$. This setup introduces a scale to the otherwise gapless system and by integrating out modes of mass $\propto {1\over \D}$ which are supported along the common longitudinal directions of the defects, we arrive at the EFT that captures the fusion limit ($i.e.$ small $\D$ expansion of the defect two-point function $\la\cD_1(0)\cD_2(\D) \ra$). The same consideration carries over when the defect worldvolume $\Sigma$ is curved as long as the curvature scale is large compared to $\D$.
For a pair of conformal boundaries on $\Sigma$ (and $\Sigma_\D$ related by a transverse displacement), the defect effective action takes the same form as in \eqref{thermaleff} \cite{Diatlyk:2024zkk}
\ie 
S_{\rm BEFT}(\D)={1\over \D^{d-1}}\int d^{d-1}x\sqrt{g} \left( -\cE +\dots \right) \,,
\label{Deff}
\fe
where the leading Wilson coefficient $\cE$ is the negative of the Casimir energy density and the subleading Wilson coefficients multiply extrinsic and intrinsic curvature invariants whose influence on the DCFT data can be found in \cite{IshibashiDefLong}. 
The similarity between \eqref{Deff} and \eqref{thermaleff} is not a coincidence. In fact the thermal EFT can be thought of as special case of the boundary EFT introduced here. This is achieved by applying the folding trick to CFT $\cT$ on the thermal geometry ${\mathbb{S}}_\B^1\times {\mathbb{S}}^{d-1}$, which produces the tensor product CFT $\cT\otimes \overline{\cT}$ where $\overline{\cT}$ is defined with the orientation reversed, on an interval of length $\B\over 2$ with boundary condition $\cB$ (and its orientation reversal) from folding the trivial interface. Therefore, the thermal EFT is equivalent to the boundary EFT with $\D={\B\over 2}$ with these distinguished boundary conditions. The inclusion of angular momentum chemical potentials in the thermal EFT can be similarly accounted for in this boundary EFT by replacing $\cB$ at one end of the interval by the boundary condition that arises from the fusion of the rotation symmetry defect of $\cT$ with $\cB$, which amounts to a twisted identification for spinning states of $\cT$ on ${\mathbb{S}}^{d-1}$.

We also derive general constraints on the Wilson coefficient $\cE$ for general conformal defects of dimension $p$ from unitarity (reflection positivity). In particular, we prove that $\cE$ is non-negative when the pair of parallel $p$-dimensional defects $\cD_1,\cD_2$ are related by orientation reversal and more generally $\cE$ satisfies a convexity-like property (see Appendix~\ref{app:genbounds}).

To see how the DEFT encodes universal data of the DCFT, it is convenient to consider the setup where the defects wrap concentric spheres, which are conformally equivalent to another setup where the defects wrap disjoint spheres of the same radius. We refer to the two conformal frames as the annulus frame and the bulk-OPE frame respectively. By tuning the dimensionless modulus in the the defect two-point function, it is clear that the annulus frame is naturally described by the DEFT while in the bulk-OPE frame, as the name suggests, involves decomposition of the individual spherical defects into bulk local operators at their centers and then summing over the two-point functions. It is via the second frame that the DCFT data such as \eqref{Bopf} naturally enters and consistency connects their asymptotic behavior at large $\Delta$ to the Wilson coefficients in the DEFT.  

In the rest of the letter, we will provide more details on the points summarized above. We point out that our analysis for conformal boundaries in $d=2$ CFTs is equivalent to studying the Cardy condition (also known as the closed-open duality) for the annulus (cylinder) partition function \cite{Cardy:1989ir,Cardy:2004hm}. The DEFT approach advocated here is its generalization to general $d$. In particular, important technical ingredients that enter in the $d=2$ Cardy condition such as Ishibashi states and annulus blocks have direct generalizations at higher $d$, some of which are already available from \cite{Nakayama:2015mva,Gadde:2016fbj} which we make use here for the boundary case and a more complete analysis is included in \cite{IshibashiDefLong}. Similar to the case of thermal EFT analyzed in \cite{Benjamin:2023qsc,Benjamin:2024kdg}, we expect the defect EFT introduced here to give access to more general defect observables, than just the defect structure constants (as in \eqref{Bopf}), such as a generalization of the $d=2$ results in \cite{Kusuki:2021gpt,Numasawa:2022cni} to higher dimensions. 

Finally, a major triumph of the Cardy formula is its prediction for the microscopic entropy of 
black hole microstates, which can be understood via the AdS$_3$/CFT$_2$ correspondence \cite{Strominger:1997eq, Hartman:2014oaa}. More generally, the statistics of heavy operators and relevant OPE data in CFT with AdS dual is captured by the on-shell actions of nontrivial saddle-points such as black hole and wormholes in the gravitational path integral \cite{Shaghoulian:2015lcn, Kraus:2016nwo, Chandra:2022bqq,Kusuki:2022wns,Chandra:2022fwi,Chandra:2024vhm}. It would be interesting to compare our findings to such bulk geometries. For top-down holographic constructions where the defects are realized by probe brane-antibrane pair in string theory, it would also be interesting to use our CFT results to learn about open-string tachyon condensation and the emission of closed string radiation \cite{Sen:2003bc,Sen:2004nf}. 

{\it Note added}: Recently we were informed that a related setup is studied and will appear in \cite{cusp}.

\section{Ishibashi States and Annulus Two-Point Function in General Dimension}
\label{sec:IshibashiStates}
We start by setting up the defect two-point function for a pair of conformal boundaries $\cB$ and its orientation reversal $\overline\cB$. 
In the annulus frame, the conformal boundaries wrap concentric spheres of radius $r_1$ and $r_2$ respectively with $r_1>r_2$. In the bulk-OPE frame, the conformal boundaries carve out two balls of size $r$ separated by distance $D$ between their centers. The two frames are related by a standard conformal transformation (see $e.g.$ \cite{PhysRevB.51.13717}) with
\ie 
r={4r_1r_2\over r_1-r_2}\,,\quad D={4\sqrt{r_1r_2}(r_1+r_2)\over r_1-r_2}\,.
\fe
We introduce the dimensionless modulus $\B$ which satisfies,
\ie 
e^{{\B\over 2}}={r_1\over r_2}\,,\quad \cosh {\B\over 4}={D\over 2r}\,,
\label{moduli}
\fe
and the normalization is chosen such that $\B$ is the standard inverse temperature for the CFT on the thermal geometry before folding into the annulus geometry (conformally equivalent to the cylinder) as mentioned in the introduction. The defect two-point function in this case is simply the annulus partition function for the boundary state $|\cB\ra$,
\ie 
Z_{\cB \overline\cB}(\B) \equiv \la \cB | e^{-{\B\over 2}H } |\cB\ra \,.
\label{aPF}
\fe
where $H$ is the radial Hamiltonian.
From \eqref{moduli}, the fusion limit corresponds to $\B \ll 1$ while the bulk OPE limit is $\B \gg 1$. As discussed in the introduction, the former is naturally described by the DEFT \eqref{Deff} with $\D={\B\over 2}$, and the latter is a sum over two-point functions of scalar operators that appear in the decomposition of the spherical boundaries.

The boundary state $|\cB\ra$ on the unit sphere ${\mathbb{S}}^{d-1}$  can be decomposed as 
\ie 
|\cB\ra=\sum_\phi   C_\phi |\phi \rrangle\,,
\label{eq:BoundaryDec}
\fe 
into Ishibashi states $|\phi \rrangle$ which are in one-to-one correspondence with scalar primary operators $\phi$ and automatically preserve the residual conformal symmetry $\mf{so}(d,1)$. The coefficient $C_\phi$ is the boundary structure constant that determines the one-point function of $\phi$ as in \eqref{Bopf} and the $2^{\Delta_\phi}$ is a consequence of the conformal transformation from the planar defect to the spherical defect.
The Ishibashi states in general $d$  were first introduced in \cite{Nakayama:2015mva} and we give an alternative simple derivation in Appendix~\ref{app:ishibashi}. The result is the following combination of $\phi$ and its scalar descendants
\ie 
|\phi \rrangle =&  \sum_{n=0}^\infty \kappa^\phi_n\LP^n\phi(0) | 0 \rangle 
\,,\quad 
 \kappa^\phi_n= \frac{ 2^{-2n}  }{   n! \left(\Delta_\phi+1-{d\over 2}\right)_{n} } \,,
 \label{ishibashi}
\fe
where $(a)_n\equiv \Gamma(a+n)/\Gamma(a)$ is the Pochhammer symbol.

Although the Ishibashi states are not normalizable, they have well-defined and diagonal matrix elements under radial evolution, which we will refer to as the annulus blocks
\ie 
&\chi^{\textrm{Annulus}}_{\Delta_\phi} (q)
\equiv \llangle \phi|e^{-{\B\over 2}H}|\phi \rrangle
\\
&=\frac{q^{ \Delta_\phi\over 2}}{(1 - q)^{d-1}}
    \phantom{0}_2F_1\left(1 - d + \Delta_\phi, 1 - \frac{d}{2}, 1-\frac{d}{2}+\Delta_\phi, q \right) \,,
    \label{annulusblock}
\fe
where $q\equiv e^{-\B}$ and the last line follows from \eqref{ishibashi} after using \eqref{eq:NormalizationDescendant}. It is easy to check that the annulus block are eigenfunctions for the differential equation 
\ie 
{4 q^{{d\over 2}+1}\over (1-q)^d} {d\over dq} \left({ (1-q)^d \over  q^{{d\over 2}-1}}{d\over dq} \right) \chi^{\textrm{Annulus}}_{\Delta_\phi}=\Delta_\phi(\Delta_\phi-d)\chi^{\textrm{Annulus}}_{\Delta_\phi}\,,
\label{annulusCAS}
\fe
which agrees with the conformal Casimir equation of \cite{Gadde:2016fbj}.

Putting the above together, we obtain the general decomposition of the annulus partition function into annulus blocks labeled by scalar primaries $\phi$,
\ie 
Z_{\cB \overline\cB}(\B)=\sum_\phi C_\phi^2  \chi^{\textrm{Annulus}}_{\Delta_\phi}(e^{-\B})\,.
\label{annulusdecomp}
\fe

\subsection*{Comparison to Thermal Blocks by Folding} 
As mentioned in the introduction, CFT on the thermal geometry ${\mathbb{S}}^1_\B\times {\mathbb{S}}^{d-1}$ is related to that on the annulus by folding. 
The thermal partition function
\ie 
Z(\B) = \int_0^\infty  d\Delta \sum_{\vec J} \rho^{\textrm{prim}}(\Delta,{\vec J}) \chi_{\Delta,\vec J}(q)\,,
\fe
decomposes into thermal blocks weighted by the density $\rho^{\textrm{prim}}(\Delta,\vec J)$ of conformal primaries with dimension $\Delta$ and spin $\vec J$. Each thermal block is a conformal character at zero angular chemical potential \cite{Dolan:2005wy} ,
 \ie
 \chi_{\Delta, \vec J} (q) = \frac{ q^{\Delta}}{(1 - q)^d} \textrm{dim}_{\vec J}  \,,
 \fe 
where $\textrm{dim}_{\vec J}$ is replaced by the ${\rm SO}(d)$ character and the denominators are modified accordingly when angular chemical potentials are turned on.  

Consistency upon folding and unitarity requires the thermal blocks to decompose into annulus blocks \eqref{annulusblock} with non-negative coefficients $a_n$,
\ie 
    \chi_{\Delta, \vec J} (q) = \textrm{dim}_{\vec J}\sum_{n = 0}^{\infty}  a_n\chi_{2\Delta+2n}^{\textrm{Annulus}} (q)\,.
    \label{torustoannulus}
\fe 
Said differently, each thermal block $\chi_{\Delta,\vec J}$ for CFT $\cT$ corresponds to an \textit{extended} annulus block of dimension $2\Delta$ for the CFT $\cT\otimes \overline{\cT}$ due to the higher spin symmetries from the extra conserved stress-energy tensor. The explicit coefficients (see Appendix~\ref{app:blocks}) are
\ie 
    a_n=\frac{1}{n!}\frac{\left(\frac{d}{2}\right)_n\left(2\Delta+n+1-d\right)_n}{\left(2\Delta+n-\frac{d}{2}\right)_n }\,,
    \label{ancoeff}
\fe 
which will be needed to translate our general universal formula for CFT with defects to the special case of CFT without defects.

\section{Asymptotic Boundary Structure Constants}
We now derive universal formula for the behavior of boundary structure constants $C_\phi$ by analyzing the fusion limit of the annulus partition function \eqref{annulusdecomp}. We define
\ie 
B(\Delta)\equiv \sum_{\phi} C_\phi^2 \D( \Delta_\phi-\Delta)\,,  
\label{eq: smallBetaPartition}
\fe
and consider $\B\ll 1$ in \eqref{annulusdecomp}. The annulus block simplifies in this limit to $\chi^{\textrm{Annulus}}_{\Delta_\phi} \to e^{-{\B\over 2}\Delta_\phi} \B^{-{d\over 2}}$ (see Appendix~\ref{app:blocks} and \eqref{highTblock} for details) and we obtain
\ie 
{1\over \B^{d\over 2}}\int_0^\infty d \Delta  B(\Delta)  e^{-{\B\over 2} \Delta}\xrightarrow[]{\B\ll 1} e^{{2^{d-1}\over \B^{d-1}}S_{d-1}\cE}\,,
\label{annulushighT}
\fe
where the RHS is the leading contribution from the DEFT \eqref{Deff} and $S_{d-1} \equiv 2 \pi^{\frac{d}{2}} /\Gamma \left( \frac{d}{2}\right)$. Note that this immediately implies that the boundary Casimir energy between $\cB$ and $\overline\cB$ is non-positive ($i.e.$ $\cE \geq  0$) due to the positivity and divergent behavior on the LHS. See Appendix~\ref{app:genbounds} for an alternative argument that holds for general $p$ dimensional conformal defects \footnote{This non-positivity property of the Casimir energy also generalizes to conformal defects of higher dimensions following the same reasoning here \cite{IshibashiDefLong}. This property for line defects is also derived independently in \cite{cusp}. After the first version of this paper was posted, we also became aware that the similar result is derived independently in \cite{petretal}.}.

Performing the inverse Laplace transform
\ie 
{1\over 4\pi i}\int_{-i\infty}^{i\infty} d\B  e^{{2^{d-1}\over \B^{d-1}}S_{d-1}\cE + {\B\over 2}\Delta} \B^{d\over 2}\,,
\label{eq: SaddlePoint}
\fe
by saddle point approximation and taking into account the one-loop contributions, we obtain the following universal formula for the weighted squared boundary structure constants at high energy,
\ie
\label{StructConstWeighted}
B(\Delta) \sim & {2^{d-1\over 2} \over \sqrt{d \pi }}{((d-1)\cE S_{d-1})^{d+1\over 2d}\over  \Delta^{{2d+1\over 2d}}}
e^{  {d \left ({\Delta\over d-1}\right)^{d-1\over d}(\cE S_{d-1})^{1\over d} }  }\,.
\fe 
As in the case of other Cardy-type formula, the above holds only after averaging over a micro-canonical energy window which is small compared to $\Delta$. These statements can be made more rigorous, including the leading corrections and precise size of the energy window, by using Tauberian theorems (see \cite{
Das:2017vej,Qiao:2017xif,Mukhametzhanov:2019pzy,Pal:2019zzr,Mukhametzhanov:2020swe,Das:2020uax}). 

For the special annulus setup related by folding the CFT $\cT$ on the thermal geometry, we have the following relations between the BEFT quantities and those in the thermal EFT,
\ie 
\cE={f_{\rm thermal}\over 2^{d-1}}\,,\quad \Delta=2\Delta_{\rm thermal}\,,
\fe
where the first equation comes from the reduction $\B \to {\B/2}$ from folding, and the second equation comes from the identification between scalar operators in the folded theory $\cT\otimes \overline{\cT}$ and general operators in $\cT$ (see also \eqref{torustoannulus}). From these relations we see immediately that the exponential behavior in \eqref{StructConstWeighted} matches onto the expected asymptotic density of primary operators in $\cT$. The one-loop factors also match after taking into account the relation between the annulus and the thermal blocks (see in particular \eqref{largedimblocksrel}).

To further isolate the asymptotic behavior of the structure constants $C_\phi$ at high energy,
we introduce the density of scalar primaries 
\ie 
\rho^{\textrm{prim}}(\Delta)\equiv \sum_\phi \D(\Delta_\phi-\Delta)\,,
\fe
which can be derived  (see Appendix \ref{derivationDensityPrim}) following the general discussion in \cite{Benjamin:2023qsc},
\ie 
\rho^{\textrm{prim}}(\Delta)
\sim 
\A_d {f^{(d+1)(d+2)\over 4d}\over \Delta^{d^3+5d+2\over 4d}}
e^{d \left ({\Delta\over d-1}\right)^{d-1\over d}(f S_{d-1})^{1\over d} }\,.
\label{densityofstatesprimeris}
\fe 
where $\A_d$ is a positive constant independent of $f$ and $
\Delta$. Therefore, the boundary one-point function of a typical heavy primary operator is
\ie  
\label{StructConstGeneral}
C^{\rm typical}_\phi \equiv {\sqrt{B(\Delta)\over \rho^{\rm prim}(\Delta)}} 
\propto  \Delta^{d^2+1\over 8} 
e^{{d\over 2} \left ({\Delta\over d-1}\right)^{d-1\over d} S_{d-1}^{1\over d} (\cE^{1\over d}- f^{1\over d}) }\,,
\fe
  where a constant $\Delta$ independent prefactor has been dropped as it gives subleading contribution to $\log C^{\rm typical}_\phi$ in the limit $\Delta \gg 1$. Note that this constant also receives contribution from the subleading Wilson coefficient suppressed by $\D^{d-1}$ in the effective action \eqref{Deff} for odd $d$. Note that the asymptotic behavior in \eqref{StructConstGeneral} crucially depends on the magnitude of $\cE$ versus $f$. It would be interesting to see if there exists a universal upper bound on $\cE$ by $f$.
  
\section{Examples}

Here we provide simple examples to  illustrate the universal formulas presented in the last section for the boundary structure constants.

\subsection{$d=2$ CFTs}
In our convention, the thermal free energy of a $d=2$ CFT of central charge $c$ is 
\ie 
f={\pi c\over 6}\,.
\fe
For a simple conformal boundary state $|\cB\ra$ of the $d=2$ CFT, the boundary Casimir energy is universal ($i.e.$  $|\cB\ra$ independent), and given by \cite{CARDY1988377}
\ie 
\cE={  \pi c \over 24} = {f\over 4}\,.
\label{2dCasimir}
\fe
This is because a strip  with the corresponding boundary condition $\cB$ and its orientation reversal on the two sides is related to a half-space with a single boundary condition $\cB$ by a conformal transformation. The ground state on the strip is the identity operator on the boundary after this conformal transformation and the Casimir energy is determined in terms of the bulk central charge $c$ via the usual Schwarzian derivative.

Therefore we have
\ie
B(\Delta) \sim & \frac{\pi  c^{3\over 4}}{2^{3\over 2}\ 3^{3\over 4} \Delta ^{5\over 4}}
e^{\pi   \sqrt{c \Delta\over 3  } } \,,
\fe 
and correspondingly
\ie 
C^{\rm typical}_\phi  \propto
\Delta^{5\over 8}e^{-{\pi\over 2}   \sqrt{c \Delta\over 3  } } \,,
\fe
which implies that the boundary structure constants decay exponentially for typical heavy operators in the bulk. 

Note that for $d=2$, annulus block in \eqref{annulusblock} coincides with the chiral thermal character for a scalar quasi-primary, for which the above formulas apply. It is straightforward to generalize to the case of Virasoro primaries, which amounts to studying the usual Cardy condition on the cylinder in the limit of small length,
\ie 
 \int_0^\infty d \Delta  B_{\rm Vir}(\Delta)  e^{-{\B\over 2} \Delta}\xrightarrow[]{\B\ll 1} \sqrt{2\pi \over  \B}e^{\pi^2 (c-1)\over 6\B }\,,
\fe
where we have used that, the Virasoro character behaves as,
\ie 
\chi^{\rm Vir}_{\Delta}(\B)={q^{{\Delta\over 2}-{c-1\over 24}}\over \eta(q)} 
\xrightarrow[]{\B\ll 1} e^{-{\B\over 2}\left({\Delta}-{c-1\over 12} \right)} {\sqrt{\B\over 2\pi}} e^{\pi^2 \over 6 \B }\,.
\fe
Consequently, from inverse Laplace transform as before, we find
\ie 
B_{\rm Vir}(\Delta)
\sim &  {1 \over 2 \sqrt{\Delta}}
e^{\pi   \sqrt{(c-1) \Delta\over 3  } }\,,
\fe 
and the boundary structure constant for a typical heavy Virasoro primary behaves as
\ie 
C^{\rm typical}_{\phi\,\rm Vir} \equiv {\sqrt{B_{\rm Vir}(\Delta)\over \rho_{\rm Vir}^{\rm prim}(\Delta)}}  \propto
\Delta^{1\over 4}e^{-{\pi\over 2}   \sqrt{(c-1) \Delta\over 3  } } \,,
\fe 
where we have used that the density of scalar Virasoro primary states is (see $e.g.$ \cite{Benjamin:2019stq})
\ie 
\rho^{\textrm{prim}}_{\rm Vir}(\Delta)
\sim 
{1 \over  2\Delta}
e^{2\pi   \sqrt{(c-1) \Delta\over 3  }}\,.
\fe 
Again, we find that the boundary structure constant for heavy Virasoro primaries decay exponentially.

\subsection{Boundaries for Free Scalars and Free Fermions} \label{sec:FreeScalar}

The relation \eqref{2dCasimir} between the Casimir energy $-\cE$ and the thermal free energy $f$ does not hold for CFTs of dimension $d>2$. Nonetheless, our universal formulas for the boundary structure constants still apply given the BEFT.

Here we discuss free theories with conformal boundary conditions. For concreteness, we focus on the theory of a real scalar and a Dirac fermion in $d=3$, though the discussion generalizes straightforwardly to such free theories in higher $d$. We will construct their boundary states $|\cB\ra$ similar to what was done for $d = 2$ in \cite{Gaberdiel:2002my}, and verify \eqref{annulushighT} by comparing with the corresponding boundary Casimir energy.

We start with the theory of a free real scalar $\Phi$ in $d=3$ on the Euclidean cylinder $\mR \times \mathbb{S}^2$ where $\tau$ labels the Euclidean time coordinate. The $\Phi$ field  has the following decomposition into eigenfunctions of the Laplacian on $\mathbb{S}^2$ with coordinates $(\theta,\varphi)$,
\ie 
 \Phi(\theta,\varphi,\tau) =\sum_{\ell=0}^\infty \sum_{m=-\ell}^{\ell}{Y_{\ell, m}(\theta,\varphi)\over \sqrt{2\omega_\ell}}\left( a_{\ell, m} e^{\omega_\ell  \tau} + a^\dagger_{\ell, m} e^{-\omega_\ell \tau}\right) \, ,
 \fe 
where $Y_{\ell, m}$ are the standard real spherical harmonics, the frequency is fixed by the equation of motion to be  $\omega_\ell\equiv \ell+{1\over 2}$, and the creation and annihilation operators obey the commutation relation $[a_{\ell,m},a^\dagger_{\ell',m'}] = \delta_{\ell,\ell'}\delta_{m,m'}$. 

The normal ordered Hamiltonian of the free scalar on $\mathbb{S}^2$ is given by 
\begin{gather}
    H = \sum_{\ell=0}^\infty \sum_{m=-\ell}^{\ell}\omega_\ell \left(a^\dagger_{\ell,m}a_{\ell,m}+\frac{1}{2}\right) \,.
    \label{scalarham}
\end{gather}
The thermal partition function follows immediately,
\ie 
    Z_\Phi (\B)\equiv \tr e^{-\beta H}  
    =\prod_{\ell =0}^{\infty}\dfrac{1}{\left(1-e^{-\beta \left(\ell +\frac{1}{2}\right)}\right)^{2\ell +1}}\,,
    \label{Zphi}
\fe 
where the contribution from zero modes  in \eqref{scalarham} vanishes by zeta function regularization.  The above expression, after taking logarithm, can be rewritten as,
\ie 
  \log Z_\Phi (\B)=-&\sum^\infty_{\ell=0} (2\ell +1) \log\left(1 - e^{-\beta\left(\ell+\frac12\right)}\right)\\
    =&\, \sum^\infty_{m=1} \frac{\sinh\left(m\beta\right)}{4m \sinh^{3}\left(m \frac{\beta}{2}\right)}
\fe 
where the second line comes from expanding the log and summing over $\ell$ first. In general dimensions, this becomes
\cite{Benjamin:2023qsc, Chang:1992fu, Iliesiu:2018fao}
\ie 
 \log Z_\Phi (\B)= &\,\sum^\infty_{m=1} \frac{\sinh\left(m\beta\right)}{2^{d-1} m \sinh^{d}\left(m \frac{\beta}{2}\right)}  \\
    =& \frac{2 \zeta(d)}{\B^{d-1}} - \frac{ (d - 4) \zeta (d - 2)}{12  \B^{d-3}} + \mathcal{O}\left(\beta^4 \right) \,,
    \label{sphereFscalar}
\fe 
where we have included the high temperature expansion in the second line and the leading term determines the negative thermal free energy density $ f_{\Phi} = \frac{2 \zeta(d)}{S_{d-1}}$ \footnote{Note that the thermal partition function is divergent at $d = 3$, as can be seen from  the order $\beta^2$ term in \eqref{sphereFscalar}. Relatedly, the free energy $ \log Z_\Phi (\B)$ contains a $\log \beta$ term in its high temperature expansion.
These arise from the gapless mode of the scalar compactified on ${\rm S}^1$ (see also \cite{Benjamin:2023qsc}). In any case, the thermal free energy $f_\Phi$ still captures the leading growth of the density of heavy states.}.

Familiar conformal boundary conditions of a free scalar are Dirichlet and Neumann boundary conditions and we denote the corresponding boundary states as $|\cB_D\ra$ and $|\cB_N\ra$ respectively. The Dirichlet boundary condition $\Phi(\theta,\varphi,\tau=0)| \cB_D\rangle = 0 $ demands that $(a_{\ell,m} + a^\dagger_{\ell,m}) |\cB_D\rangle  =0$ for all $\ell,m$, similarly the Neumann boundary condition $\partial_\tau\Phi(\theta,\varphi,\tau=0)| \cB_N\rangle = 0$ is solved by $(a_{\ell,m}- a^\dagger_{\ell,m}) |\cB_N\rangle  =0$. The boundary states on $\mathbb{S}^2$ at $\tau=0$ are then fixed 
\ie 
    |\cB_D\rangle =& \, g_D\left( \prod^\infty_{\ell=0} \prod_{m=-\ell}^\ell e^{-\frac12 a^\dagger_{\ell,m} a^\dagger_{\ell,m}} \right) |0\rangle \, ,
    \\
        |\cB_N\rangle = & \,g_N\left( \prod^\infty_{\ell=0} \prod_{m=-\ell}^\ell e^{\frac12 a^\dagger_{\ell,m} a^\dagger_{\ell,m}} \right) |0\rangle \, ,
    \label{Dbs}
\fe 
up to overall constants $g_D,g_N$ which coincide  with the boundary $g$-function for even $d$ and are scheme dependent for odd $d$. These expressions are direct generalizations for the boundary states of the non-compact free boson in two dimensions \cite{Gaberdiel:2002my}. 

The cylinder partition function with two such Dirichlet boundary states separated by a distance $\B/2$ is 
\ie
    Z_{\cB_D\cB_D}(\beta)= \langle \cB_D | e^{-\frac{\beta}{2} H}  |\cB_D\rangle = g_D^2 \sqrt{Z_\Phi(\B)}\,,
    \label{cylinderD}
\fe
where we have used \eqref{Dbs} explicitly and compared to \eqref{Zphi}  in the last equality.
Similarly for a pair of Neumann boundary states, the cylinder partition function reads
\ie 
Z_{\cB_N\cB_N}(\beta)= \langle \cB_N | e^{-\frac{\beta}{2} H}  |\cB_N\rangle =  g_N^2 \sqrt{Z_{\Phi}(\B) }\,,
    \label{cylinderN}\fe
and the mixed partition function vanishes $Z_{\cB_D\cB_N}(\beta)=0$.
These are consistent with folding the free scalar theory on ${ \mathbb{S}}_\B^1\times {\mathbb{S}}^2$ which implies 
\ie 
Z_\Phi(\B)= Z_{\cB_D\cB_D}(\beta) Z_{\cB_N\cB_N}(\beta)\,,
\label{foldfreescalar}
\fe
as long as $g_N g_D=1$.
The boundary state for the doubled theory from folding is  a direct sum of Neumann and Dirichlet boundary states for two real scalars  and the cylinder partition function factorizes accordingly as in \eqref{foldfreescalar}.

From the above, we deduce that the corresponding boundary Casimir energies are related to the thermal free energy of the $d=3$ real scalar by (recall $\D={\B\over 2}$ in \eqref{Deff})
\ie 
\cE_D=\cE_N={f_\Phi\over 8}\,.
\fe
It is straightforward to repeat this analysis for general $d$ and obtain
\ie 
\cE_D=\cE_N={f_\Phi\over 2^d}={ \zeta(d)\over 2^{d-1}S_{d-1}}\,,
\fe
in agreement with previous results  (see for instance \cite{Diehl:2011sy, Diatlyk:2024zkk}).
 
We now discuss boundary states for the theory of a Dirac fermion $\Psi$ in $d=3$. Since it is conceptually similar to the free scalar case described above, we will outline the main results here and relegate the details to Appendix~\ref{app:FreeFermion}. 

We denote the two sets of creation and annihilation operators from the decomposition of fermion field  $\Psi(\theta,\varphi,\tau)$ as $b^\dagger_{j,m},b_{j,m}$ and $c^\dagger_{j,m},c_{j,m}$ with positive half-integer $j$ and half-integer $m$ satisfying $|m|\leq j$. 
The two boundary states $|\cB_\pm\ra$ corresponding to the conformal boundary conditions $\sigma^3 \Psi = \pm \Psi$ at $\tau=0$ are
\ie 
|\cB_\pm \rangle = g_\pm \prod\limits_{j=\frac{1}{2}}^{\infty}\prod \limits_{m=-j}^{j}\left(1 \mp i c^\dagger_{j,m} b^\dagger_{j,m}\right)|0\rangle\,,
\fe 
with constant coefficients $g_\pm$.

 The thermal partition function of the free fermion is
\ie \label{eq: PartitionFFreeFermion}
    Z_\Psi (\beta) = \tr e^{-\beta H} = \prod_{j=\frac12}^\infty \left(1+e^{-\beta (j+\frac12)}\right)^{4j+2} \,,
\fe
whose logarithm can be rewritten as,
\ie 
\log     Z_\Psi (\beta)=&\sum_{m=1}^\infty  \frac{(-1)^{m+1}}{m \sinh^2 \frac{m \beta}{2}} 
\\
=& \frac{3 \zeta(3)}{\B^2}  - \frac{\log 2 }{3} + \mathcal{O}\left(\beta^4\right)  
\,,
\fe
and we have included its high temperature expansion in the second line. The leading contribution determines the thermal free energy which is negative of $f_\Psi={3\zeta(3)\over 4\pi}$ in $d=3$.  
Once again the cylinder partition functions are related to thermal partition function by
\ie 
Z_{\cB_+\cB_+}(\beta) =  g_+^2 \sqrt{Z_{\Psi}(\B) }\,,\quad 
Z_{\cB_-\cB_-}(\beta) =  g_-^2 \sqrt{Z_{\Psi}(\B) }\,,
\label{cylinderp}\fe
while $Z_{\cB_+\cB_-}(\beta) =0$. Consequently, we conclude as before that
\ie 
\cE_+=\cE_-={f_\Psi\over 8}={3\zeta(3)\over 32\pi}\,.
\fe
It is easy to see that these relations generalizes for arbitrary $d$ to 
\ie 
\cE_+=\cE_-={f_\Psi\over 2^d}=2^{\lfloor\frac{d}{2}
\rfloor+1}\left(1-2^{1-d}\right){\zeta(d)\over 2^d S_{d-1}}\,.
\fe
where $\lfloor\frac{d}{2}
\rfloor$ is the greatest integer less than or equal to $\frac{d}{2}$, and we have used $f_\Psi=2^{\lfloor\frac{d}{2}
\rfloor+1}\left(1-2^{1-d}\right){\zeta(d)\over S_{d-1}}$ \cite{Chang:1992fu}.
These results for the boundary Casimir energy are in agreement with previous results in  
\cite{DePaola:1999im, Diatlyk:2024zkk}.

 \section*{Acknowledgements}

We thank Nathan Benjamin, Gabriel Cuomo, Tom Hartman, Zohar Komargodski and Sridip Pal for helpful questions and discussions. We also thank Petr Kravchuk, Alex Radcliffe and Ritam Sinha for sharing with us their draft after the first version of this paper appeared which has some overlaps. 
The work of YW was
supported in part by the NSF grant PHY-2210420 and by the Simons Junior Faculty Fellows program.
F.K.P.
is currently a Simons Junior Fellow at New York University and supported
by a grant 855325FP from the Simons Foundation.

\newpage
\onecolumngrid

\appendix
\section{Ishibashi states} \label{app:ishibashi}
In this section we show how a codimension-one spherical conformal defect $\cD$ (boundary or self-interface) may be represented as a sum of local operators inserted at its center. We have used this decomposition in \eqref{eq:BoundaryDec} for a boundary state in the main text. To start, let us consider a spherical codimension-one conformal defect of radius $r$ centered at the origin. The one-point function of a scalar primary operator in the presence of this defect is given by \cite{Billo:2016cpy} 
\ie
    \langle \phi(x) \rangle_\cD = \frac{ C_\phi r^{\Delta_\phi}}{|x^2 - r^2|^{\Delta_\phi}}\,,
\label{eqApp:OnePtphi}
\fe
which in the near defect limit ($i.e$  $|x|\to r$), behaves as
\ie 
\langle \phi(x) \rangle_\cD \to \frac{C_\phi}{2^{\Delta_{\phi}} ||x|-r|^{\Delta_\phi}}\,,
\fe 
which matches onto \eqref{Bopf} for the flat defect.
Now we decompose the defect state 
\ie
    |\cD \rangle  = \sum_{\phi} C_\phi | \phi \rrangle \,,
    \label{Ddecomp}
\fe 
in terms of boundary Ishibashi states $| \phi \rrangle$ associated to each scalar primary $\phi$ in the CFT and defined by a linear combination of the primary and its scalar descendants,
\ie 
 | \phi \rrangle\equiv  \sum_{\phi} C_\phi F_{\Delta_{\phi}}(\Box,r) \phi(0) | 0 \rangle \, ,
 \label{phiishi}
\fe
where we have introduced the function $F_{\Delta_{\phi}}(\Box,r)$ where $\Box$ denotes the scalar Laplacian on $\mR^d$ to collectively account for the descendants. 

These boundary Ishibashi states in general spacetime dimension have been discussed already in \cite{Nakayama:2015mva} and below we provide an alternative way to derive them. Generalizations of such Ishibashi states for conformal defects of higher codimensions can be found in \cite{IshibashiDefLong}. To fix the function $F_{{\Delta_\phi}}(\Box,r)$, we note that by taking the inner product of both sides of \eqref{phiishi} with $\la \phi(x)|$, consistency with \eqref{eqApp:OnePtphi} and \eqref{Ddecomp} requires that 
\ie 
     F_{{\Delta_\phi}}(\Box,r)  \left[\frac{1}{x^{2{\Delta_\phi}}}\right] = \frac{ r^{\Delta_\phi}}{(x^2 - r^2)^{\Delta_\phi}} \, .
\fe 
This completely fixes $F_{{\Delta_\phi}}(\Box,r)$ to be,
\begin{gather}
    F_{\Delta_\phi}(\Box,r) = (r )^{{\Delta_\phi}} \phantom{|}_0F_1\left(1-\frac{d}{2}+{\Delta_\phi},\frac{r^2 \Box}{4}\right) \, ,
    \label{eq:IshibashiBoundaries}  
\end{gather}
where we have made use of the following identity 
\begin{equation}
    \Box^{n}  \left( \frac{1}{x^{2\Delta_{\phi}}} \right) = \frac{4^{n}\left(\Delta_{\phi}\right)_n\left(1-\frac{d}{2}+\Delta_{\phi}\right)_n}{x^{2\Delta_{\phi}+2n}}   \, . 
\end{equation}
Expanding the differential operator in \eqref{eq:IshibashiBoundaries},
we obtain the series expansion for the Ishibashi state of a scalar primary $\phi$,
\ie
     |\phi \rrangle = \sum_{n=0}^\infty r^{\Delta_{\phi} + 2 n}\kappa^\phi_n\LP^n\phi(0) | 0 \rangle 
\,, \hspace{0.5cm} \kappa^\phi_n= \frac{ 2^{-2n}  }{   n! \left(\Delta_\phi+1-{d\over 2}\right)_{n} } \,, 
\label{ishidecomp}
\fe 
This agrees with what was found in \cite{Nakayama:2015mva}. Such boundary Ishibashi states provide a generalization of the more familiar two dimensional Ishibashi states \cite{1989MPLA....4..251I} to higher spacetime dimensions. 

As a further check of this decomposition we may look at the descendant one-point functions, which are given by 
\begin{equation}
    \langle  \cD |  (P^2)^n \phi (0) | 0 \rangle   = C_{\phi} \kappa^{\phi}_n  \langle \phi | [(K^2)^n, (P^2)^n] | \phi \rangle \equiv  C_{\phi} \kappa^{\phi}_n \mathcal{N}_{n} \, , 
    \label{Nndef}
\end{equation}
where we have defined the normalization $\mathcal{N}_{n}$ of descendant states $(P^2)^n|\phi\ra$ by the last equality. Here $P_\m,K_\n$ are standard translation and special conformal generators that obey the commutation relation $[P_\m,K_\n]=2(\D_{\m\n} D + M_{\m\n})$.
Alternatively we may calculate the one-point function of descendants by just taking derivatives of the primary one-point function. Equating the two, we obtain the following identity,
\begin{equation}
    C_{\phi} \kappa^{\phi}_n \mathcal{N}_{n} = \Box^n \frac{C_{\phi}}{(1 - x^2)^{\Delta_{\phi}}} \bigg|_{x^2 \rightarrow 0}
    \label{eq:SanityCheckIshibashi}
\end{equation}
where we have set the radius of the spherical defect to $r=1$. Using the following commutator
\begin{equation}
    [K^2, (P^2)^n] | \phi \rangle = 4 n (2 \Delta_{\phi} + 2 n - d) [2n+d-2] [\Delta_\phi+n-1](P^2)^{n - 1} | \phi \rangle \, ,
\end{equation}
it is easy to see that for generic $n$, we have from the second expression in \eqref{Nndef},
\begin{equation}
    \kappa^{\phi}_n\mathcal{N}_{n} =4^n \left(\frac{d}{2}\right)_n \left(\Delta_\phi\right)_n\, .
    \label{eq:NormalizationDescendant}
\end{equation}
which matches the derivative on the right hand side in \eqref{eq:SanityCheckIshibashi}, hence providing an additional check of our decomposition \eqref{ishidecomp}.

\section{Annulus conformal blocks}
\label{app:blocks}

In this section, we spell out the relation between thermal and annulus blocks as we discussed in Section~\ref{sec:IshibashiStates} and discuss their high temperature limit which is used in deriving the universal formula for boundary structure constants of primary operators at high energy.

To find the positive coefficients $a_n$ in \eqref{torustoannulus} which relate the annulus and the thermal blocks, we make use of the eigenvalue equation \eqref{annulusCAS} satisfied by the annulus blocks. This implies the following orthogonality condition,
\begin{gather}
   \frac{1}{2\pi i} \oint_C \frac{dq}{q^2} (1-q)^d q^{1-\frac{d}{2}}\chi_{\Delta_\phi}^{\textrm{Annulus}} (q)\chi_{d-\Delta_\phi'}^{\textrm{Annulus}} (q)=\delta_{\Delta_\phi,\Delta_\phi'}\,, \quad \quad \Delta_\phi-\Delta_\phi'=2\mathbb{Z}\,,
\end{gather}
where the contour $C$ circles around $q=0$ counterclockwise, and the normalization is  obtained
by a Laurent expansion of the integrand. The condition $\Delta_\phi-\Delta_\phi'=2\mathbb{Z}$ makes sure that there are no branch cuts and hence the total derivative term integrates to zero. Multiplying both sides of \eqref{torustoannulus} by annulus block with dimension $d - 2\Delta_\phi - 2 n$ and appropriate powers of $q$, we can find coefficients $a_n$ explicitly by the contour integral,
\begin{gather}
    a_n=\frac{1}{2\pi i} \oint_C \frac{dq}{q^2}q^{\Delta}q^{1-\frac{d}{2}}\chi_{d-2\Delta-2n}^{\textrm{Annulus}} (q)=\frac{1}{n!}\frac{\left(\frac{d}{2}\right)_n\left(2\Delta+n+1-d\right)_n}{\left(2\Delta+n-\frac{d}{2}\right)_n }\,.
    \label{a_ncoeff}
\end{gather}

Next, we analyze the behavior of annulus blocks in the limit of small  width and large conformal dimensions, $i.e$ $\beta \ll 1$ and $\Delta_\phi \gg 1$ that we use to study \eqref{annulushighT}. We are interested in the regime where $\beta^d \Delta_\phi \sim 1$ as suggested by the saddle point in \eqref{eq: SaddlePoint}. To proceed, we write \eqref{annulusblock} as
\ie 
\chi_{\Delta_\phi}^{\textrm{Annulus}} (q)
   =e^{-\frac{\beta}{2}\Delta_\phi}\sum \limits_{n=0}^{+\infty}\frac{\left(\frac{d}{2}\right)_{n}\left(\Delta_\phi\right)_{n}}{\left(1-\frac{d}{2}+\Delta_\phi\right)_{n}}\frac{e^{-\beta n}}{n!} \,,
\fe 
which is more suitable for the limit of interest. By first expanding the coefficients in front of $e^{-\beta n}$ for large $\Delta_\phi$ and then summing over $n$, we obtain the following,
\ie 
  \chi_{\Delta_\phi}^{\textrm{Annulus}} (q) 
     =\frac{e^{-\frac{\beta }{2}\Delta_\phi}}{(1-e^{-\beta})^{d/2}}\left (1+\frac{d(d-2)e^{-\beta}}{4(1-e^{-\beta})\Delta_\phi}+\mathcal{O}\left(\frac{1}{(1-e^{-\beta})^2\Delta_\phi^2}\right)\right)\, ,
     \label{highTblock}
\fe 
where the subleading terms are of the form $(\beta \Delta_\phi)^{-n}$ and are suppressed in the high temperature limit with $\beta^d \Delta_\phi \sim 1$. 

As a consistency check, let us compare the high temperature expansion of the annulus and the thermal blocks using 
\eqref{torustoannulus}. In fact, for this comparison, we can take $\B$ fixed and large $\Delta_\phi$. Indeed,
using that  
\ie 
\lim \limits_{\Delta_\phi \to \infty}a_n=\frac{1}{n!}\left(\frac{d}{2}\right)_n\,,
\fe 
we recover the thermal block as below,
\ie 
     \lim_{\Delta_\phi \to \infty}\sum_{n = 0}^{\infty}  a_n\chi_{2\Delta_\phi+2n,d}^{\textrm{Annulus}} (e^{-\beta})= \sum_{n=0}^{+\infty}\frac{1}{n!}\left(\frac{d}{2}\right)_n\frac{e^{-\beta \Delta_\phi-\beta n}}{\left(1-e^{-\beta}\right)^{\frac{d}{2}}}= \frac{e^{-\beta \Delta_\phi}}{\left(1-e^{-\beta}\right)^{d}} \, .
     \label{largedimblocksrel}
     \fe

\section{Derivation of the density of scalar primaries}\label{derivationDensityPrim}
In this section we derive the asymptotic density of heavy scalar primaries \eqref{densityofstatesprimeris}. 
Following the general discussion in \cite{Benjamin:2023qsc},  the thermal partition function $   Z(\B)$ is determined at high temperature by the thermal EFT, which to leading order is only sensitive to the thermal free energy $f$,
\ie 
   Z(\B)=  \int d \Delta \sum_{J_i}  \rho_{d}^{\rm prim}(\Delta,\vec J)\frac{e^{-\beta \Delta} \chi_{\vec J}(\beta \Omega_i)}{\left(1-e^{-\beta}\right)^{\epsilon}\prod\limits_{i=1}^{n}\left(1-e^{-\beta(1+i \Omega_i)}\right)\left(1-e^{-\beta(1-i \Omega_i)}\right)} \xrightarrow{\B\ll 1} \exp\left[\frac{ fS_{d-1} }{\beta^{d-1}\prod\limits_{i=1}^{n}\left(1+\Omega^2_i\right)}\right]
     \,.
    \label{partitionDdens}
\fe 
Here and below  $n\equiv [\frac{d}{2}]$ is the rank of $\mf{so}(d)$.
For convenience, we also introduce 
$\epsilon$ such that $\epsilon=0$  for even $d$ and $\epsilon=1$ for odd $d$. An irreducible $\mf{so}(d)$ representation is denoted as $\vec J\equiv (J_1,\dots,J_n)$ where $J_i\in \mZ_{\geq 0}$ label the charges of the  highest weight state with respect to rotations in the $n$ two-planes in $\mR^d$.  The corresponding angular chemical potentials are $\B \Omega_i$. In \eqref{partitionDdens} we have included the contributions from the conformal descendants by the denominators in the second expression. To isolate operators of a given spin, we will also need the $\mf{so}(d)$ character, which can be found in \cite{10.1214/aop/1176988864},
\ie 
     \chi_{\vec J}(\beta \Omega_s)=\frac{e^{i\beta \sum_{s=1}^n \Omega_s (J_s+ j_s)}+{\rm Weyl~reflections}}{\left(\prod \limits_{1\leq r\leq n}\left[2i\sin{\frac{\beta \Omega_r}{2}}\right]\right)^{\epsilon}\prod \limits_{1\leq s<r\leq n}\left[2\left(\cos{\beta \Omega_r}-\cos{\beta \Omega_s}\right)\right]}\,, \quad j_s\equiv \frac{\epsilon}{2}+s-1\,,
\fe 
where we have suppressed additional terms in the numerator that come from Weyl reflections (with appropriate signs).

To determine the asymptotic density $\rho^{\rm prim}_d(\Delta,\Vec{J})$ of primary states of spin $\vec{J}$, we take the inverse Laplace transform of the last expression in \eqref{partitionDdens}. Here we focus on the scalar density, which is given by
\ie 
    \rho_{d}^{\rm prim}(\Delta,\Vec{0})=\int \limits_{-i\infty}^{+i\infty} \frac{d\beta}{2\pi i}\int \limits_{\gamma-i\infty}^{\gamma+i\infty}\prod_{k=1}^{n}\frac{d\Omega_k}{2\pi i}e^{S\left(\beta,\Omega_s\right)}g_{d}(\beta,\Omega_i)\,, 
    \label{integraldensity}
\fe 
with 
\ie 
    S\left(\beta,\Omega_s\right)=\frac{ f S_{d-1} }{\beta^{d-1}\prod\limits_{i=1}^{n}\left(1+\Omega^2_i\right)}+\beta \Delta-i\beta \sum \limits_{s=1}^{n}\Omega_s j_s\,,
    \label{ILTS}
\fe 
and 
\ie 
    g_d(\beta,\Omega_i)  
    =(i \beta)^n (1-e^{-\beta})^{\epsilon}\prod \limits_{1\leq r\leq n}\left(2i\sin{\frac{\beta \Omega_r}{2}}\right)^{\epsilon}\prod \limits_{1\leq s<r\leq n} 2\left(\cos{\beta \Omega_r}-\cos{\beta \Omega_s}\right)  \prod\limits_{i=1}^{n}\left(1-e^{-\beta(1+i \Omega_i)}\right)\left(1-e^{-\beta(1-i \Omega_i)}\right)\,,
    \label{gdfactor}
\fe 
where the $(i \beta)^n$ factor is the Jacobian of $\delta^{(n)}\left(i\beta (\vec{J}-\vec{J}')\right)$ from integrating over the angular chemical potentials $\Omega_k$. 

We proceed to analyze the inverse Laplace transform \eqref{integraldensity} by saddle point approximation which is valid at large $\Delta$. It is convenient to treat for the moment $j_s$ in \eqref{ILTS} as a formal variable and replace $\B \Omega_r \to i {\partial \over \partial j_r}$ in \eqref{gdfactor}. Consequently, we can focus on  \eqref{ILTS} to analyze the saddle point and its contribution to \eqref{integraldensity} (the perturbative corrections in $\Delta$ will receive contributions from the Hessian and also \eqref{gdfactor} by taking derivatives in $j_r$).

The saddle point equations of the action $S\left(\beta,\Omega_s\right)$ were studied in \cite{Benjamin:2023qsc}.  At large $\Delta$, the solution is given by
\ie 
\B_*=\left({f  S_{d-1}(d-1)\over \Delta}\right)^{\frac{1}{d}}\,,\quad 
\Omega_{k*}=-i \frac{(d-1)j_k}{2\Delta} \,.
\label{saddlesol}
\fe
and the saddle point value of $S\left(\beta,\Omega_i\right)$ is
\ie 
 S (\beta_*,\Omega_{i*}) ={d f S_{d-1}\over \B_*^{d-1}}  - {\B_*^{d+1}\over 2 f S_{d-1}} \sum_{k=1}^n j_k^2 +\dots  \,,
 \label{Ssaddle}
\fe  
up to subleading contributions (we have kept the leading dependence on $j_k$ which we will use later to evaluate the one-loop corrections).

To determine the one-loop (perturbative) contribution to the inverse Laplace transform in \eqref{integraldensity}, we expand around the saddle $\Omega_i=\Omega_{i*}+\eta_i,\beta=\beta_{*}+\eta_{n+1}$ and denote the Hessian matrix by $S''_{ab}$ with $a,b=1,\dots,n+1$,
\ie 
 \int\prod_{c=1}^{n+1} \frac{d\eta_c}{2\pi i } e^{{1\over 2} \sum_{a,b=1}^{n+1} 
 \eta_a S''_{ab}(\beta_*,\Omega_{i*})\eta_b}
 \propto 
 \frac{1}{i^n} f^{-\frac{n-1}{2d}}\Delta^{-\frac{(d-1)n+d+1}{2d}}
 \,.
 \label{hessian}
\fe
This integral is evaluated 
by choosing steepest descent contour for $\eta_c$ which is along the real axis for $c=1,...,n$ and the imaginary axis for $c=n+1$. Combined with the fact that the Hessian matrix $S''_{ab}$ is block diagonal in the large $\Delta$ limit, with top left block behaving as $-f\B_*^{1-d}\D_{ij}$ and bottom right block (single entry) behaving as $f\B_*^{-1-d}$, the result on the RHS of \eqref{hessian} follows (up to an overall real $f$-independent constant whose precise value is not important here). 

The additional one-loop contribution comes from the $g_d(\B,\Omega_i)$ factor \eqref{gdfactor} by taking derivatives of \eqref{Ssaddle} with respect to $j_k$. This is equivalent to evaluating $g_d(\B,\Omega_i)$ on the saddle \eqref{saddlesol} which gives (up to an overall real $f$-independent constant)
\ie 
g_d(\B_*,\Omega_{i*}) \propto  i^n f^{\frac{n (n+\epsilon +2)+\epsilon }{d}} \Delta ^{-\frac{n (d (n+\epsilon -1)+n+\epsilon +2)+\epsilon }{d}}\,.
\label{oneloopgd}
\fe
Putting together \eqref{Ssaddle},\eqref{hessian} and \eqref{oneloopgd}, we obtain the final result for the asymptotic density of scalar primaries, including the one-loop contribution,
\ie 
   \rho_{d}^{\rm prim}(\Delta,\Vec{0})\sim \alpha_d\frac{f^{\frac{(d+1)(d+2)}{4d}}}{\Delta^{\frac{d^3+5d+2}{4d}}}\exp\left[d \left(\frac{\Delta}{d-1}\right)^{\frac{d-1}{d}}\left(  f  S_{d-1}\right)^{\frac{1}{d}}\right]\,,
\fe
where $\A_d$ is a positive constant independent of $f$ and $\Delta$.

\section{Free fermion on $\mathbb{R}\times\mathbb{S}^{2}  $} \label{app:FreeFermion}
In this section we discuss the Dirac fermion on a three dimensional Euclidean cylinder to set the notation right and to provide details on what we discuss in section \ref{sec:FreeScalar}. The Dirac action has the following form 
\begin{gather}
S=\int\limits_{\mathbb{R}\times\mathbb{S}^{2}  }d^3 x \sqrt{g}\Bar{\Psi} \slashed{\nabla}\Psi\,, 
    \label{action}
\end{gather}
where $\slashed{\nabla}$ is the Dirac operator on curved space $\mathbb{R}\times\mathbb{S}^{2}$ (see $e.g.$ \cite{Camporesi:1995fb,Abrikosov:2002jr}),
\begin{gather}
\slashed{\nabla}=\sigma^{1}\left(\partial_{\theta}+\frac{\cot{\theta}}{2}\right)+\sigma^{2}\left(\frac{1}{\sin{\theta}}\partial_{\phi}\right)+\sigma^{3}\partial_{\tau} \equiv \slashed{\nabla}_{S^2}+\sigma^{3}\partial_{\tau}\,,
\end{gather}
where $\sigma^i$ are Pauli matrices, and we have defined the Dirac operator on $\mathbb{S}^{2}$ which will be useful in the following. 

The Hamiltonian for the Dirac fermion on $\mathbb{S}^{2}$ is 
\begin{gather}
    H=-\int \limits_{\mathbb{S}^{2}}\Bar{\Psi} \slashed{\nabla}_{\mathbb{S}^{2}} \Psi\,, 
\end{gather}
where $\Bar{\Psi}=\Psi^{\dagger}\sigma^3$. To quantize the theory on $\mR \times \mathbb{S}^{2}$ we first need to solve the equation of motion for $\Psi$,
\ie 
\slashed{\nabla}_{\mathbb{S}^{2}}\Psi=-\sigma^{3}\partial_{\tau}\Psi\,,
\label{equaofmot}
\fe 
which can be done using the spectral decomposition of the Dirac operator on $\mathbb{S}^{2}$,
\ie 
    \slashed{\nabla}_{\mathbb{S}^{2}} \Upsilon_{j m}^{\pm}=\pm i\omega_j \Upsilon_{j m}^{\pm}\,, \quad \quad \omega_j=j+\frac{1}{2}\,,
\fe 
where $j\in \mZ_{\geq 0}+{1\over 2}$ and $m=-j,1-j,\dots,j-1,j$. The explicit form of $\Upsilon_{j m}^{\pm}$ can be found for example in \cite{Abrikosov:2002jr}. Here we only need the following properties
\begin{gather} \label{UpsilonProperties}
    \int \limits_{\mathbb{S}^{2}} \left(\Upsilon_{j'm'}^{\epsilon'}\right)^{\dagger}\Upsilon_{jm}^{\epsilon}=\delta_{\epsilon, \epsilon'}\delta_{j,j'} \delta_{m,m'}\,, \quad \quad \Upsilon_{j m}^{\pm}=\pm i \sigma^3 \Upsilon_{j m}^{\mp}\,,
\end{gather}
where $\epsilon, \epsilon'=\pm$. Consequently a basis of solutions to \eqref{equaofmot} may be constructed as
\begin{gather}
    \Psi^{\pm}_{j,m}=e^{\pm \omega_{j}\tau}\mathcal{V}_{j m}^{\pm}\,, \quad \quad \mathcal{V}_{j m}^{\pm}=\frac{\Upsilon_{j m}^{+}\pm\Upsilon_{j m}^{-}}{\sqrt{2}}\,.
\end{gather}
To proceed , we perform canonical quantization by expanding the Dirac fermion in this basis multiplied by creation operators $b^\dagger_{j,m},c^\dagger_{j,m}$ and annihilation operators $b_{j,m},c_{j,m}$, 
\ie 
    \Psi(\theta,\phi,\tau) = &\sum_{j,m} \left[ c_{j,m}  e^{-\omega_j \tau} \mathcal{V}^-_{j m}(\theta,\phi)+ b^\dagger_{j,m} e^{\omega_j \tau} \mathcal{V}^+_{j m}(\theta,\phi)\right] \, , \\
       \Psi^{\dagger}(\theta,\phi,\tau) = &\sum_{j,m} \left[ c^{\dagger}_{j,m}  e^{\omega_j \tau} \left(\mathcal{V}^-_{j m}(\theta,\phi)\right)^{\dagger}+ b_{j,m} e^{-\omega_j \tau} \left(\mathcal{V}^+_{j m}(\theta,\phi)\right)^{\dagger}\right] \, , 
       \label{fielddecomposdermions}
       \fe 
and impose canonical anti-commutation relations,
       \ie 
    \{ b_{j, m}, b^{\dagger}_{j', m'}\} =  \{c_{j, m}, c^{\dagger}_{j', m'}\} = \delta_{j,j'}\delta_{m,m'}\,.
   \fe 
Note that Hermitian conjugation on the Euclidean cylinder involves a reflection in time $\tau \rightarrow -\tau$. Using this decomposition \eqref{fielddecomposdermions}, the normal ordered Hamiltonian becomes
\ie 
    H=\sum_{j,m} \omega_j \left(c^\dagger_{j,m} c_{j,m}+b^\dagger_{j,m} b_{j,m}-1\right) \, .
    \fe
The thermal partition function immediately follows from the following trace over the Hilbert space,
    \ie 
    Z_\Psi(\beta) = \tr e^{-\beta H}  = \prod_{j=\frac12}^\infty \left(1+e^{-\beta (j+1/2)}\right)^{4j+2} \, .
\fe 
where as in the case of scalars, zero mode contribution from $H$ to $Z(\beta)$ vanish by zeta function regularization. 

\section{General bounds on Casimir energy and  one-point function}
\label{app:genbounds}
 In this section, taking inspiration from \cite{Bachas:2006ti} and using unitarity (reflection positivity), we derive general constraints on the Casimir energy $-\cE$ (defined in a similar way as in \eqref{Deff}) for a pair of parallel $p$-dimensional conformal defects $\cD_1$ and $\cD_2$ (see $e.g.$ \cite{Diatlyk:2024zkk} for a general discussion). In addition, we also derive a strict upper bound for the one-point function coefficients of bulk local operators in the presence of a boundary. 
 
 Let us start by considering two parallel conformal defects $\cD_i$ and $\overline\cD_j$ of dimension $p$ in flat space separated by a transverse distance $z$. Then as discussed in previous sections (see also \cite{Diatlyk:2024zkk}), 
 the extensive piece in the defect volume $V$ is controlled by the Casimir energy $-\cE_{ij}$ as below,
 \ie 
 \lim_{V/z^p\to \infty } {1\over V} {\log\, \langle \cD_i(0) \overline\cD_j(z) \rangle } = {\cE_{ij}\over z^p} \,.
 \fe
 In the following we will work in this limit.

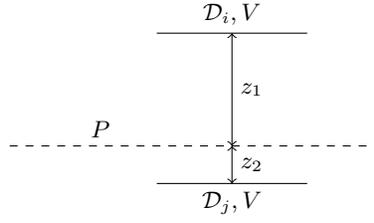
\begin{figure}[!htb]
     \centering
     \begin{tikzpicture}
        \draw (-1,0)--(1,0) node[midway,below] {$\cD_j,V$};;
        \draw[<->] (0,0.5)--(0,2) node [midway, right] {$z_1$};
        \draw[dashed] (2,0.5)--(-3,0.5) node [near end,above] {$P$};
        \draw[<->] (0,0.5)--(0,0) node [midway, right] {$z_2$};
        \draw (-1,2)--(1,2) node[midway,above] {$\cD_i,V$};
     \end{tikzpicture}
     \caption{Configuration of two conformal defects of dimension $p$ parallel to a plane $P$, the first defect is separated from plane $P$ by a distance $z_1$ and the second is separated by a distance $z_2$.}
     \label{fig:config}
 \end{figure}

 Let us consider a plane parallel to the defects  and located at a distance $z_1$ from the first defect and $z_2$ from the second defect (see Figure~\ref{fig:config}). Then, taking $z$ to be the Euclidean time direction, the defect correlator computes the overlap of the (time-evolved) defect states, we thus have the following Cauchy-Schwartz inequality
 \begin{gather}
     \left|\langle \cD_i(z_1)|  \cD_j(z_2) \rangle \right|^2 \leq\langle \cD_i(z_1)|  \cD_i(z_1) \rangle  \langle \cD_j(z_2)|  \cD_j(z_2) \rangle\,.
 \end{gather}
 
 Writing out explicitly these overlaps we get the following inequality for $\cE_{ij}$ (see also \cite{Bachas:2006ti})
 \ie 
     \frac{2\cE_{ij}}{(z_1+z_2)^{p} } \leq  \frac{\cE_{ii} }{(2 z_1)^{p}} + \frac{\cE_{jj} }{(2 z_2)^{p}} \,,
 \fe 
 after taking the limit $V\to \infty$ and keeping $z_{1,2}$ fixed.

In the special case where the defects are related by orientation reversal ($i.e.$ $i=j$) the above becomes
\begin{gather}
     \frac{2\cE_{ii}}{(z_1+z_2)^{p} } \leq  \frac{\cE_{ii}}{(2 z_1)^{p}} + \frac{\cE_{ii}}{(2 z_2)^{p}}\,,
\end{gather} 
Using the convexity of the function $\frac{1}{z^p}$ we conclude 
that this inequality could be satisfied only if $-\cE_{ii} \leq 0$ leading to the non-positivity of the Casimir energy as stated around \eqref{annulushighT}.
 
Now we derive an upper bound for the general $\cE_{ij}$ in the form of a convexity-like constraint. For that we rescale $z_1+z_2=1$ and set $z_1 = t,z_2 = 1- t$, leading to
 \begin{gather}
     \cE_{ij} \leq \frac{\cE_{ii} }{2^{p+1} t^{p}} + \frac{\cE_{jj} }{2^{p+1} (1-t)^{p}}\,,\quad \forall t \in \left[0,1\right] \, .
 \end{gather}
After minimizing the RHS of the above equation, we obtain the following inequality \footnote{We would like to thank Petr Kravchuk, Alex Radcliffe and Ritam Sinha for sharing their draft, where similar (but weaker) inequalities were derived independently \cite{petretal}.}
 \begin{gather}
     \min_{t\in[0,1]} \left(\frac{\cE_{ii}}{2^{p+1} t^{p}} + \frac{\cE_{jj}}{2^{p+1} (1-t)^{p}}\right) =  \left(\frac12 \cE_{ii}^\frac{1}{p+1} + \frac12 \cE_{jj}^\frac{1}{p+1}\right)^{p+1} \geq \cE_{ij} \,, 
\end{gather}
where the strongest bound is achieved at 
\ie 
t=t_{*}\equiv \frac{\cE^{\frac{1}{p+1}}_{ii}}{\cE^{\frac{1}{p+1}}_{ii}+\cE^{\frac{1}{p+1}}_{jj}}\,.
\fe 
Note that this inequality is saturated only when $\cD_i=\cD_j$.

This method can also be used to derive a general bound for the one-point function of bulk primary operators in the presence of a conformal defect. For illustration, here we restrict to the case of a conformal boundary. We consider a cylinder geometry $\mathbb{R}_\tau \times \mathbb{S}^{d-1}$ and a boundary state $\cB$ that is located at time $\tau=-\tau_1$ and a primary scalar state (corresponding to operator $\phi$) at time $\tau=\tau_2$. We have the following Cauchy-Schwartz inequality as before,
 \begin{gather}
 \left| \langle \cB(-\tau_1)| \phi(\tau_2) \rangle \right|^2 \leq  \langle \cB(-\tau_1)| \cB(\tau_1) \rangle  \langle \phi(-\tau_2)|\phi(\tau_2) \rangle \,,
 \end{gather}
 or more explicitly the following inequality on the one-point function coefficient $C_\phi$,
 \begin{gather}
    \frac{1}{2^{2\Delta_\phi}}\frac{C_\phi^2}{\sinh^{2\Delta_\phi}(\tau_1+\tau_2)} \leq e^{E(2\tau_1)} \frac{1}{2^{2\Delta_\phi}\sinh^{2\Delta_\phi}(\tau_2)} \Rightarrow   C^2_\phi \leq \min_{\tau_{1,2}} \left[ e^{E(2\tau_1)} \left(\frac{\sinh(\tau_1+\tau_2) }{\sinh(\tau_2)}\right)^{2\Delta_\phi} \right]\,,
 \end{gather}
where we have introduced $E(\tau_1+\tau_2)\equiv  \log \langle \cB(-\tau_1)| \cB(\tau_2) \rangle$. Note that the RHS as a function of $\tau_2$ is monotonically decreasing for all $\tau_1$. Therefore we reduce the minimization problem to
 \begin{gather}
     2 \log \left[C_\phi\right] \leq  
      -\max_{\tau_1} \left[\left(-\Delta_\phi \right) \left(2\tau_1\right) - E(2\tau_1)  \right] 
     =
     -\tilde{E}(-\Delta_\phi)\,,
 \end{gather}
 where $\tilde{E}(\cdot )$ denotes the Legendre transform of the function $E(\cdot )$. Note that this inequality is  never saturated. Indeed, if we had an equality, from Cauchy inequality  we must conclude that $|\cB \rangle \propto |\phi \rangle$, which is impossible. Thus, this inequality must be strict,
 \begin{gather}
     2 \log \left[C_\phi\right] < -\tilde{E}(-\Delta_\phi)\,.
 \end{gather}

\bibliographystyle{apsrev4-2}
\bibliography{defect}
\end{document}